\documentclass{article}      % use "amsart" instead of "article" for AMSLaTeX format
\usepackage{graphicx}               % Use pdf, png, jpg, or epsÂ§ with pdflatex; use eps in DVI mode
\usepackage{caption}
\usepackage{subcaption}
\usepackage{sidecap}
\usepackage{floatrow}
\usepackage{amsmath}
\usepackage{amsthm}
\usepackage{amsfonts}
\usepackage{verbatim}
\usepackage{bm}
\usepackage{nicefrac}

\usepackage[numbers, sort&compress]{natbib}
\usepackage{pdfpages}

\usepackage[final]{neurips_2020}

\newcommand{\expect}{\mathbb{E}}

\newcommand{\poissondist}{\textrm{Pois}}
\newcommand{\ud}{\textrm{d}}

\title{Adaptation Properties Allow Identification of Optimized Neural Codes}
\author{
Luke Rast \\
Harvard University \\
lukerast@g.harvard.edu
\And 
Jan Drugowitsch\\
Department of Neurobiology \\
Harvard Medical School \\
jan\_drugowitsch@hms.harvard.edu
}

\begin{document}

\maketitle

\begin{abstract}
The adaptation of neural codes to the statistics of their environment is well captured by efficient coding approaches. Here we solve an inverse problem: characterizing the objective and constraint functions that efficient codes appear to be optimal for, on the basis of how they adapt to different stimulus distributions. We formulate a general efficient coding problem, with flexible objective and constraint functions and minimal parametric assumptions. Solving special cases of this model, we provide solutions to broad classes of Fisher information-based efficient coding problems, generalizing a wide range of previous results. We show that different objective function types impose qualitatively different adaptation behaviors, while constraints enforce characteristic deviations from classic efficient coding signatures. Despite interaction between these effects, clear signatures emerge for both unconstrained optimization problems and information-maximizing objective functions. Asking for a fixed-point of the neural code adaptation, we find an objective-independent characterization of constraints on the neural code. We use this result to propose an experimental paradigm that can characterize both the objective and constraint functions that an observed code appears to be optimized for.

\end{abstract}

\section{Introduction}

It is well known that neural codes adapt to the distribution of stimulus inputs that they receive on a variety of timescales.
For example, in early fly vision, in addition to being well adapted to unchanging statistics of natural images \cite{Laughlin1981}, neural sensitivity adapts to both fast- and medium- timescale changes in the distribution of stimuli \cite{Brenner2000,Fairhall2000,Fairhall2001}.
Similar effects have been observed in mammalian vision \cite{Smirnakis1997,Dan1996}.
Such adaptation effects are often viewed through the lens of the efficient coding hypothesis \cite{Attneave1954,Barlow1961}, which states that neural codes are well-adapted to their task and environmental contexts, and so can be modeled as solutions to task and environment dependent optimization problems.
This is a reasonable modeling approach since we expect the result of an adaptation process to be stable in the sense of a first-order optimality condition: there are no directions that the code will continue to adapt along once it has come to equilibrium. 

Several recent works in efficient coding have focused on optimizing the local Fisher information in the neural code.
Starting from mutual information approximating objective functions and independent Poisson neural populations \cite{Brunel1998}, this approach has been generalized to accommodate a variety of objective functions \cite{Ganguli2010,Ganguli2014,Wang2012,Wang2016NeuralComp} and metabolic constraints \cite{Ganguli2010,Ganguli2014,Wang2016NIPS} under various parametrized neural activity models.
In an activity distribution agnostic formulation, these Fisher efficient codes have also been shown to predict general behavioral effects for various objective functions \cite{Wei2012,Wei2015,Wei2017,Morais2018}.
This diversity of result leads to hope that we can fully characterize the influence of objectives, constraints, and activity noise distributions in Fisher efficient codes and, in so doing, describe observed neural codes in terms of the optimization problem that they appear to be solving.
Similar recent approaches have fit objective functions to observed behavior \cite{Wu2018,Daptardar2019} and neural activity \cite{Chalk2019} by inverse reinforcement learning/optimal control, although this often leads to underspecified problems.

In this work we show that, for Fisher information based efficient codes with one-dimensional stimulus variables, the objective and constraint functions that a code is optimized for can be fully characterized by examining how the neural code adapts to changes in the stimulus distribution.
Working with flexible objective functions and constraints in a minimally parameterized setting, we first provide solutions to broad classes of Fisher efficient coding problems, generalizing previous solutions.
We show that constraints enforce deviation from classical histogram equalization behaviors, while changing the objective results in qualitative changes to the adaptation behavior of the neural code.
These solutions allow us to derive tests for the presence of constraints and for mutual information approximating objectives.
Next, asking for a stationary-point of the neural code adaptation, we find an objective independent characterization of constraints on the neural code.
Using this result, we propose an experimental paradigm to characterize both objective functions and constraints based on the observed adaptation properties of neural codes.

\section{Results}
Let $s$ be a one-dimensional stimulus of interest, drawn from a distribution, $p_s(s)$ as specified by a task or environment.
This stimulus is encoded in the activity $\bm{r}$ of one or more neurons in a population according to an encoding model $p(\bm{r} | s)$.
In an efficient code, this encoding has been optimized for performance under a specific objective while obeying particular constraints imposed on neural activity.
Specification of an efficient coding problem thus requires choosing \cite{Park2017}: (i)~what features of the code can be optimized, (ii)~how the performance of the code is measured, and (iii)~what constraints are imposed on neural activity.
Here, we solve in general a class of efficient coding problems with: optimized sufficient statistics, Fisher information-based objective functions, and activity-dependent constraints.

\subsection{Approach}
\begin{figure}[htb!]
    \centering
    \includegraphics{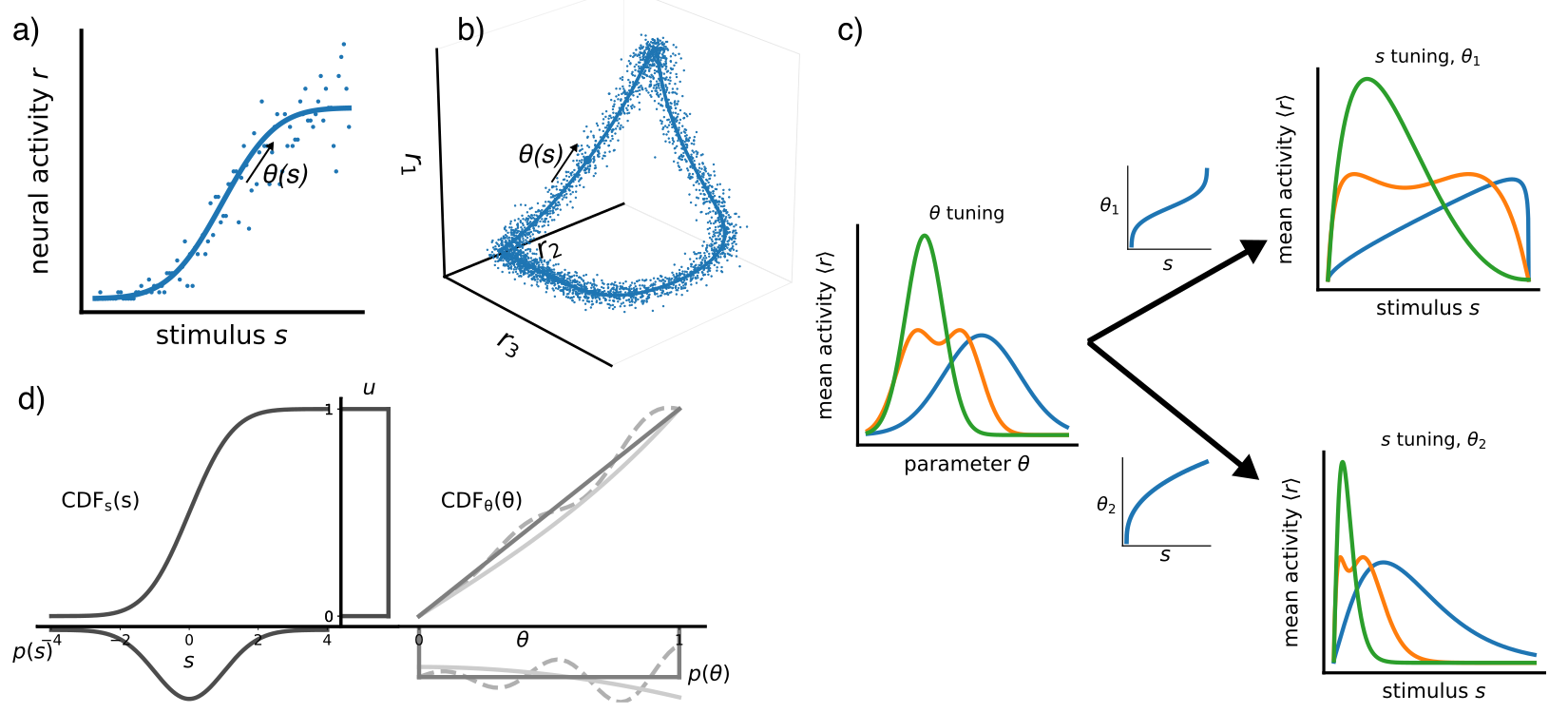}
    \caption{Our efficient coding setup.
    (a) \& (b) Example encoding models, $p(\bm r| \theta)$, for  (a) single neuron and (b) neural population.
    In (a), $\theta$ specifies the mean activity (tuning) of neurons, while noise is be assumed Poisson. 
    In (b), the encoding model specifies a particular one-dimensional submanifold of neural activity (blue line) that the stimulus is coded along. 
    (c)~Effect of different $\theta(s)$ functions on the tuning of neurons in example (b).
    Functions $\theta(s)$ determine the "speed" that we move along the one-dimensional manifold, but not the relative average activity among different neurons (colors).
    (d)~Due to monotonicity, the $\theta(s)$ mapping can be characterized in terms of cumulative distributions, $\theta (s) = \textrm{CDF}^{-1}_\theta(\textrm{ CDF}_s(s))$, or probability densities $\ud \theta / \ud s = p_s(s) / p_\theta(\theta)$, illustrated here are different $\theta(s)$ mappings (line style) for the same $p_s(s)$.
    }\label{fig:cartoon}
\end{figure}

Our approach makes three key assumptions.
First, we assume that the one-dimensional stimulus of interest is coded noisily around a one-dimensional submanifold of neural activity.
The submanifold itself is held fixed while we optimize how different stimuli are associated with different locations in this neural submanifold.
This assumption is made implicitly in a wide range of previous work \cite{Brunel1998,Ganguli2010, Ganguli2014,Wang2012, Wang2016NeuralComp, Wang2016NIPS, Wei2012, Wei2015, Wei2017, Morais2018}.
Formally, take $\theta(s)$ to be a \textit{sufficient statistic} for the stimulus in the neural activity,
\begin{equation}
p(\bm r | s ) = p(\bm r | \theta(s)) , \label{eq:assumeSS}
\end{equation}
so that $\theta(s)$ captures everything about stimulus $s$ that is encoded in neural activity $\bm r$.
We assume that this $\theta(s)$ is \textit{continuous} and \textit{strictly increasing} and that $\theta$ spans a \textit{finite interval}.
This deterministic mapping $\theta(s)$ will be optimized, while the corresponding neural activity distribution $p(\bm r|\theta)$ is fixed but left unspecified, so that the results apply to any such encoding.
For example, for a single neuron with Poisson activity, we could choose $p(r | \theta) = \poissondist \left( r | \theta \right)$ in order to optimize the neuron's (monotonic) tuning curve, $\theta(s)$ (Fig.~\ref{fig:cartoon}a). 
Alternatively, in a neural population with bell-shaped turning curves, $p(\bm r | \theta)$  could specify the relative mean activities across different neurons, as well as their individual noise, while $\theta(s)$ determines how the neurons cover the stimulus space (Figs. \ref{fig:cartoon}b \& c). 
Irrespective of the encoding model's details, the $s \to \theta \to \bm{r}$ separation splits it into two steps.
The fixed $\theta \to \bm{r}$ relationship parameterizes the neural activity with a one-dimensional parameter, $\theta$, while the optimized $s \to \theta$ relationship specifies how much of stimulus space is associated with each interval of parameter space (e.g., Fig.~\ref{fig:cartoon}c).
Our results apply to any encoding model that splits in this way.

Our second and third assumptions concern the form of the objective and constraint functions that the neural code is optimized for.
We assume that the objective depends on the \textit{Fisher information}, or local sensitivity of the code, while constraints depend on the level of \textit{neural activity}.
Formally, the code optimizes
\begin{equation}
\max \; \expect_{p_s(s)} \left[  f \left( \sqrt{I_s(s)} \right)  \right] \quad \mathrm{s.t.} \quad \expect_{p_s(s)p(\boldsymbol{r}|s)} \left[ h(\boldsymbol{r}) \right] \le M , \label{eq:assumeObj}
\end{equation}
where $\expect_{p(\cdot)}$ is the expectation with respect to $p(\cdot)$.
That is, both sensitivity and activity are averaged across stimuli, while activity is additionally averaged over neural noise.
The Fisher information, defined as $I_s(s) = \expect_{p \left( \bm{r} | s \right)} \left[ \left( \frac{\partial}{\partial s} \log p \left( \bm{r} | s \right) \right)^2 \right]$, measures how well neural activity $\bm{r}$ allows us to discriminate stimulus values close to $s$ \cite{Paradiso1988,Seung1993}.
In Eq.~(\ref{eq:assumeObj}), both the local sensitivity and the local activity are wrapped in functions, $f(\sqrt{I_s})$ and $h(\bm r)$, which specify respectively how good or bad additional information or activity are.
We assume that  $f(\cdot)$ is \textit{increasing} (i.e. more information is better), and $h(\cdot)$ is \textit{non-negative} (i.e. constrained resources can be used up, but not produced by neural activity).
$f(\cdot)$ is taken to be a function of the square root of Fisher information (rather than Fisher information itself) for notational convenience: we show in the supplementary material that $f(\cdot)$ must be \emph{concave} and sub-linear at infinity to have well-behaved solutions.
As long as the objective and constraint functions satisfy these properties and are fixed when optimizing the neural code, they can otherwise be arbitrary (continuous).
We will treat $h(\cdot)$ as one-dimensional, for example, $h \left( \bm{r} \right) = \bm{1}^T \bm{r}$, which puts an upper bound on the average total activity of the neurons.
However, the results can easily be extended to higher-dimensional $h$ (i.e. multiple constraints) (see supplementary material).

Combining our assumptions, the overall problem becomes the following optimization of $\theta(s)$:
\begin{equation}
\max_{\theta(s)} \; \expect_{p_s(s)} \left[ f  \Bigg(  \sqrt{I_\theta(\theta)} \frac{\ud \theta}{\ud s} \Bigg) \right]
\quad \mathrm{s.t.} \quad 
\expect_{p_s(s)} \left[ C(\theta) \right] \le M , \quad \frac{\ud \theta}{\ud s} > 0, 
\quad \int \frac{\ud \theta}{\ud s} \ud s = L_\theta.  \label{eq:usedObj}
\end{equation}
The last two constraints express our assumptions on $\theta(s)$ in the encoding model.
In this formulation, neural activity dependence has been re-expressed in terms of the activity parameter through $p(\bm r | \theta)$. 
This encoding model enters the optimization both through $I_\theta(\theta)$, which expresses the intrinsic sensitivity of neural activity $\bm{r}$ to changes in parameter $\theta$, and through $C(\theta)$, which denotes the average level of constraint usage for parameter $\theta$.
These functions inherit their properties from $p(\bm r|s)$ and $h(\bm r)$ (in the case of $C(\theta)$) and thus are continuous and positive, but otherwise left fixed but arbitrary.
This formulation provides an appealing level of generality as it reveals principles of efficient codes irrespective of their specific neural instantiation.
However, it also means that the results do not make direct predictions about neural activity, since this additionally requires specifying or identifying an encoding model $p(\bm{r} | \theta)$.
Overall, we aim to find the optimal parameter mapping $\theta(s)$ that solves Eq.~(\ref{eq:usedObj}). This variational optimization problem can be solved by Euler-Lagrange or the Pontryagin Maximality Principle in standard fashion \cite{liberzon2011calculus,Todorov2007}.
A central goal of this work will be to ask what effect the choices of $p_s(\cdot)$, $I_\theta(\cdot)$, $f(\cdot)$ and $C(\cdot)$ have on the optimal solution, under forgiving assumptions about their forms. See supplementary material for a full list of assumptions and for the solution method.
We start with analytic solutions for relevant special cases, below.

\subsection{Special-case analytic solutions}
Here we present three special-case solutions that generalize previously studied Fisher efficient coding problems \cite{Brunel1998, Ganguli2010, Ganguli2014, Wang2012, Wang2016NeuralComp, Wang2016NIPS, Wei2012, Wei2015, Wei2017, Morais2018}.
Solutions to our efficient coding problem are functions $\theta(s)$, mapping stimuli to corresponding neural activity parameters.
As $\theta(s)$ is continuous and monotonic, this mapping can also be characterized in terms of probability densities: sampling across input stimuli, ${p_s(s) \ud s = p_\theta(\theta) \ud \theta}$, or in terms of cumulative distribution functions: $\theta (s) = \textrm{CDF}^{-1}_\theta(\textrm{ CDF}_s(s)) $ (Fig.~\ref{fig:cartoon}d).
The cumulative distribution functions of $s$ and $\theta$ cannot always be expressed in closed form, so we characterize $\theta(s)$ by its probability density function, $p_\theta(\theta)$, from which we can recover $\theta(s)$ by (if necessary, numerical) integration.
As $\theta$ determines neural activity through $p(\bm{r} | \theta)$, the $p_\theta(\theta)$ distribution tells us the weight that the efficient coding solution assigns to each part of the neural activity space; $\theta$'s with high $p_\theta(\theta)$ imply that the associated $\bm{r}$'s are more likely to occur in an optimized code, and thus preferred.

\subsubsection{Log-objective function, $f(\sqrt{I}) \propto \log I$}
\begin{figure}[htb!]
    \centering
    \includegraphics[width=\linewidth]{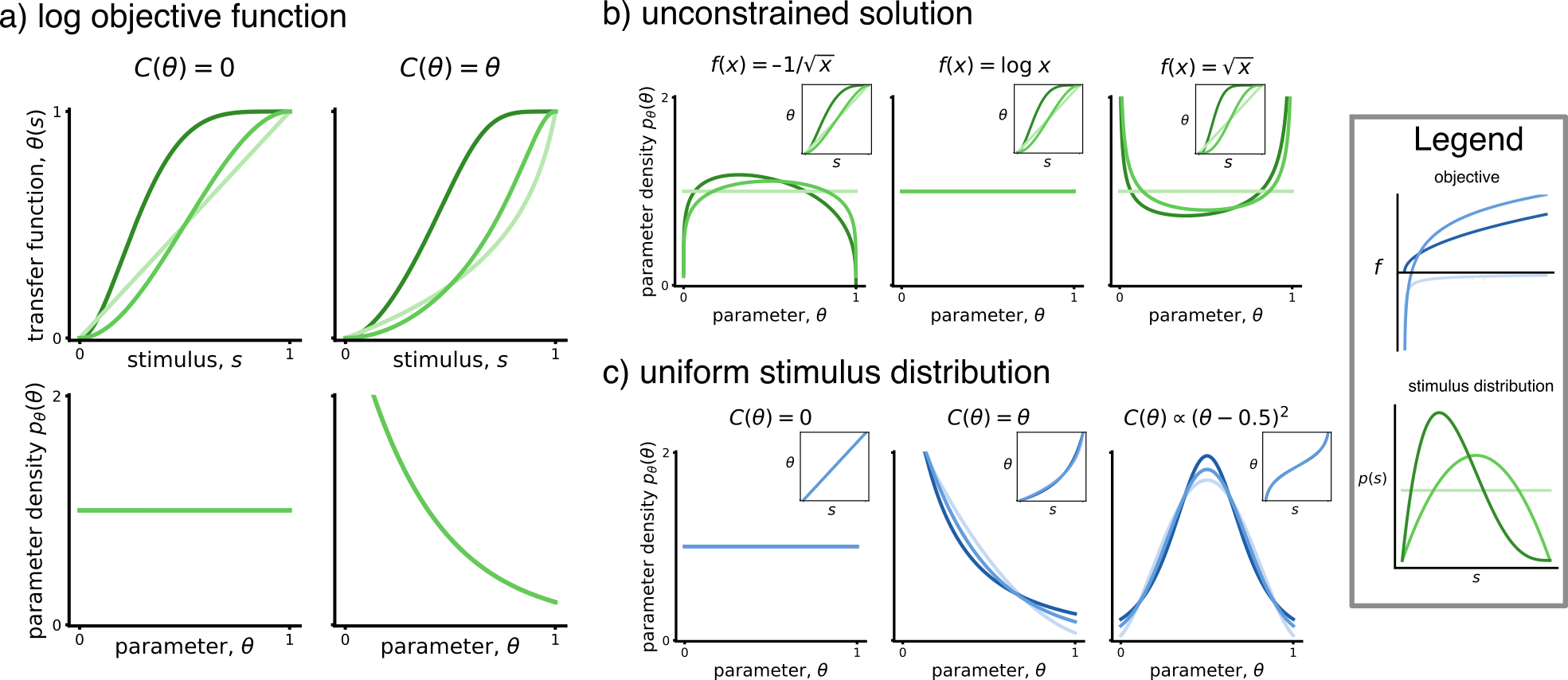}
    \caption{ Special case efficient coding solutions, $p_\theta(\theta)$.
    The corresponding $\theta(s)$ curves are shown in the top row of (a), and insets in (b) and (c).
    (a) log-objective function solution under multiple constraints (columns) and stimulus distributions (colors).
    (b) unconstrained solution under multiple objective functions (columns) and stimulus distributions (colors).
    (c) uniform stimulus distribution solution under multiple constraints (columns) and objective functions (colors).
    Plotted in the legend, the objective function can be $-1/\sqrt{x}$ (light) $\log x$ (medium), $\sqrt{x}$ (dark), and the stimulus pdf: $\textrm{Beta}(1,1)$ (light),  $\textrm{Beta}(2,2)$ (medium),  $\textrm{Beta}(2,5)$ (dark). Constraints are either $C(\theta)=\theta$ or $C(\theta) = 7 (\theta - 0.5)^2$.  For constrained panels, $M=0.3$. For all plots, $I_\theta(\theta) = 1$.}
        \label{fig:analyticResults}
\end{figure}

If $f(\sqrt I)\propto \log I$, then the objective approximates mutual information in high signal-to-noise settings \cite{Brunel1998,Wei2016}.
This objective function has been well studied in a variety of encoding models \cite{Brunel1998,Ganguli2014,Wei2012, Wei2017}, notably \cite{Wang2016NIPS}.
In our formulation, these solutions can be generalized to the optimal coding solution (Fig.~\ref{fig:analyticResults}a), characterized by:
\begin{equation}
p_\theta(\theta) = Z^{-1} \sqrt{I_\theta(\theta)} \exp(-\lambda C(\theta)), \label{eq:logpdf}
\end{equation}
an exponential family distribution with normalization factor $Z$, base measure $\sqrt{I_\theta(\theta)}$, parameter $\lambda$, and sufficient statistic $C(\theta)$.
In relation to the efficient coding problem, Eq.~(\ref{eq:usedObj}), the $M$-dependent Lagrange multiplier $\lambda > 0$ is in charge of enforcing the constraint on $C(\theta)$, while the normalization, $Z$, ensures that the $\theta$ interval has the desired length $L_\theta$.
In the absence of constraints (e.g., if $M \to \infty$), the optimal solution is determined by only the base measure, $p_\theta(\theta) \propto \sqrt{I_\theta(\theta)}$.
Otherwise, $p_\theta(\theta)$ will be as close as possible (in the Kullback-Leiber sense) to this unconstrained solution while still satisfying the constraint.
Either way, the optimal solution results in the Fisher information in neural activity $\bm{r}$ about stimulus $s$ given by
\begin{equation}
I_s(s) \propto p_s(s)^2 \exp(2\lambda C(\theta(s)) ) , \label{eq:logIs}
\end{equation}
which is larger for more likely stimuli and for stronger constraints.
We will return the properties of the found solution in Sections~\ref{sec:histogramEq} and \ref{sec:warping}.

\subsubsection{Unconstrained optimization, $M \to \infty$}
\label{sec:unconstrained}
Next, we generalize the power-law objective function analyses of \cite{Ganguli2010, Ganguli2014, Wang2012, Wang2016NeuralComp}, to examine the effect of the objective function in the absence of constraints.
For our family of well-behaved objective functions, the optimal unconstrained solution (Fig.~\ref{fig:analyticResults}b) is given by
\begin{equation}
p_\theta(\theta) = \sqrt{I_\theta(\theta)} \frac{p_s(s(\theta))} {f'^{-1} \left(\frac {Z}{p_s(s(\theta))} \right) }.  \label{eq:C0pdf}
\end{equation}
Here $f'^{-1}(\cdot)$ will exist due to concavity of $f$, while $Z$ remains the length-enforcing normalization factor.
This solution is implicit due to its dependence on $s(\theta)$.
It nonetheless shows that $p_\theta(\theta)$ follows a warped version of stimulus distribution, $p_s(s)$, where the warping is determined by $f'^{-1}(\cdot)$. Different warpings can result in substantially different adaptation behaviors of the neural code to changing stimulus distribution.
For example, for power-law objective function: $f(\sqrt{I}) \sim \alpha I^\alpha$ (which agree with previous solutions \cite{Wang2012,Wang2016NeuralComp}),
$\alpha \downarrow -\infty$ leads to ${ p_\theta(\theta) \to \sqrt{I_\theta(\theta)} p(s) }$ (a $\sqrt{I_\theta}$-scaled copy of stimulus distribution) while $\alpha \uparrow 1$ (at which point the solution becomes ill behaved) gives $p_\theta(\theta) \to \sqrt{I_\theta(\theta)} p(s)^{-\infty}$ (strongly repelled from encoding likely stimuli).
The log case occurs between these limits at $\alpha = 0$: $p_\theta(\theta) \propto \sqrt{I_\theta(\theta)} p(s)^{0}$ recovers the non-adapting parameter distribution of Eq.~(\ref{eq:logpdf}) (when unconstrained). 
Irrespective of the specific choice for objective function, the stimulus Fisher information of the unconstrained optimal solution satisfies
\begin{equation}
    f'( \sqrt{I_s(s)} ) \propto \frac{1}{p_s(s)}. \label{eq:C0Is}
\end{equation}
We will discuss further properties of this solution in Sec.~\ref{sec:warping}.

\subsubsection{Uniform probability density, $p_s(s) \propto 1$} \label{sec:uniform}
Setting the stimulus distribution to be uniform allows us to find an analytical solution that illustrates how objective and constraint functions interact.
Solving the optimization while assuming $p_s(s)  = p_s$ yields (Fig.~\ref{fig:analyticResults}c), 
\begin{equation}
p_\theta(\theta) = \sqrt{I_\theta(\theta)} \frac{p_s}{\hat f^{-1}( \lambda C(\theta) + Z  ) }. \label{eq:uniformpdf}
\end{equation}
Here, $\lambda$ and $Z$ are again the Lagrange multipliers in charge of enforcing constraint and normalization respectively, and $\hat f(\cdot)$ is the Legendre transform of $f(\cdot)$, defined as ${\hat f(x) = f'(x) x - f(x) }$, which exists uniquely because $f(\cdot)$ is concave.
As for the log-objective function case, $p_\theta(\theta)$ is as close to $\sqrt{I_\theta(\theta)}$ as possible while satisfying the constraint.
However, here the constraint is enforced through the inverse Legendre transform of the objective function.
Equivalently, closeness is measured by a more general f-divergence \cite{reid2011information}, which, in our case, is specified by the objective function: the $-f(\cdot)$ divergence.
Setting $f(x) = \log(x)$ recovers the KL divergence, and because ${\hat{f}^{-1}(x) = \exp(-x-1)}$, it also recovers the solution for the log-objective function,  Eq.~(\ref{eq:logpdf}).
Irrespective of the specific choice of object function, the stimulus Fisher information of the optimal solution satisfies
\begin{equation}
I_s(s) \propto \left( \hat f^{-1}( \lambda C(\theta(s)) + Z  ) \right)^2 . \label{eq:uniformIs}
\end{equation}
This solution is discussed further in Sec.~\ref{sec:histogramEq}.

\subsection{Reparameterization to intrinsic flat-Fisher parameters}
\label{sec:Ieffect}
None of the above solutions required us to specify the parameter-conditioned neural activity distribution, $p(\bm r | \theta)$, which comes into play in two places: it determines how we interpret parameter constraints in terms of activity, $C(\theta) = \expect_{p(\bm r | \theta)} [ h(\bm r )]$, and it determines the amount of information $I_\theta(\theta)$ that neural activity provides about $\theta$.
In all of our special-case solutions, $\smash{ \sqrt{I_\theta(\theta)} }$ scales the distribution of parameters (Eqs.~(\ref{eq:logpdf}), (\ref{eq:C0pdf}) \& (\ref{eq:uniformpdf})), but does not impact the stimulus Fisher information (Eqs.~(\ref{eq:logIs}), (\ref{eq:C0Is}) \& (\ref{eq:uniformIs})).
In other words, the emphasis on different parameter values $\theta$ is re-weighted in proportion to how intrinsically informative the associated neural responses are.
This is a general feature: $\smash{ \sqrt{I_\theta(\theta)} }$ acts as the local measure of the parameter space, as could be expected from information geometry \cite{Amari2001,amari2016information}.
Assuming that $I_\theta(\theta)$ is strictly positive (rather than only non-negative), we can introduce a reparameterization $\theta \rightarrow \hat \theta$ satisfying $\ud \hat{\theta} = \sqrt{I_\theta(\theta)} \ud \theta$.
This results in constant Fisher information, $\smash{I_{\hat{\theta}}(\hat{\theta}) \propto 1}$, so we refer to  $\smash{ \hat{\theta}}$ as the \emph{flat-Fisher space}.

This reparametrization does not affect the problem that we are solving or any features of the neural response.
After all, the parameter $\theta$ is inherited from our parameterization of the encoding model $p( \bm{r} | \theta)$, but is not unique: any invertible function of the parameter can provide a new parameterization of the problem, with its own parameter Fisher information and corresponding optimal parameter distribution.
However, by reparameterizing to the flat-Fisher parameter space, we remove the dependence on parameter Fisher information, $I_\theta(\theta)$, from
the previously derived solutions (Eqs.~(\ref{eq:logpdf}), (\ref{eq:C0pdf}) \& (\ref{eq:uniformpdf})), giving ${  p_{\hat{\theta}}(\hat{\theta}) \propto p_\theta(\theta(\hat \theta)) / \sqrt{I_\theta(\theta)}  }$.
More generally, in the formulation of the optimization problem, Eq.~(\ref{eq:usedObj}), the flat space reparameterization removes explicit $I_\theta$ dependence from the objective and renders the parameter length constraint $L_\theta$ identical to a constraint on the total available stimulus Fisher information, as previously employed by \cite{Wei2012,Wei2015, Wei2017} (see supplementary material).
Thus, by performing the reparameterization, we have removed features of the solution that result from our initial choice of parameter.
In a geometric sense, where $\theta$ parameterizes a curve through neural activity space, $\smash{\hat \theta}$ is the \emph{arc-length} parameterization of this curve: it tells us where we are on the curve as a function of how far (in the Fisher metric) we have already traveled.
The results in this flat-Fisher parameterization are intrinsic to the neural code: Solving in terms of the stimulus Fisher information we find
\begin{equation}
     p_{\hat \theta} \left( \hat \theta(s) \right) \propto p_s(s) / \sqrt{I_s(s)}.  \label{eq:flatSpace}
\end{equation}
Thus, the flat-Fisher space parameter distribution can be expressed in terms of measurable quantities: the stimulus Fisher information (discrimination performance) and stimulus distribution, and so can in principle be measured.
In the following sections, we will work in this flat-Fisher space by default, deriving tests that can be performed by measuring the flat space parameter distribution $\smash{ p_{\hat{\theta}} ( \hat{\theta} )}$, and showing how to fit it from simulated neural data.

\subsection{Signatures of objective and constraint functions in efficient codes}

Using the special case solutions expressed in the flat-Fisher space, we can derive tests for (i)~whether the neural code is shaped by constraints on neural activity, and  (ii) whether the neural code optimizes a log-objective function, as well as (iii) an experimental paradigm to fit both objective and constraint functions based on how the neural code adapts to changes in the stimulus distribution.
These tests will rely on the ability to probe the code using stimuli drawn from specific stimulus distributions.
In order to do this we must make additional assumptions, namely that the neural code \emph{adapts rapidly} to changes in the stimulus distribution $p_s(s)$ (as some neural codes have previously been shown to \cite{Smirnakis1997,Fairhall2001}) and does so by optimizing the stimulus-to-parameter mapping $\theta(s)$, while the neural activity distribution $p(\bm{r} | \theta)$ is \emph{fixed} within the timescale of the experiments.
That is to say, we assume experimental control over $p_s(s)$, and stimulus-distribution-invariant neural encoding $p(\bm{r} | \theta)$, such that only $\theta(s)$ changes when adapting to different stimulus distributions.

\subsubsection{Activity constraints enforce deviation from flat-Fisher space equalization}
\label{sec:histogramEq}
Our first test leverages deviations from a classic efficient coding signature: \emph{histogram equalization} \cite{Laughlin1981}, which refers to the observation that some measured parameter of the neural activity (often the mean firing rate) is uniformly distributed when sampling across stimuli.
Here we focus on histogram equalization behavior for the flat-Fisher parameter. Specifically, we say that 'flat-Fisher space histogram equalization' occurs when the $\smash{\hat \theta}$ distribution is uniform:
\begin{equation}
    p_{\hat \theta}(\hat \theta) \propto 1.
\end{equation}
Expressed in terms of the original parameter, this distribution has a familiar form, $\smash{ p_\theta(\theta) \propto \sqrt{I_\theta(\theta)} }$.
This is the optimal parameter distribution for both log-objective functions (Eq.~(\ref{eq:logpdf})) and uniform stimulus distributions (Eq.~(\ref{eq:uniformpdf})) when there are no constraints on activity (i.e. $\lambda = 0$).
Focusing on the uniform stimulus distribution case, we see that, regardless of the objective function that is being optimized, if the code is adapted to a uniform distribution of stimuli and there are no activity constraints, then histogram equalization will be achieved in the flat-Fisher space.
Vice versa, if neural activity constraints are binding, then the efficient coding solution will deviate from such flat space equalization when adapted to a uniform stimulus distribution.
The solutions are as close to flat space histogram equalization as possible while still satisfying the imposed constraint, with closeness measured by the objective function dependent $-f$-divergence.
This constraint enforced deviation from histogram equalization has been noted in previous works with log objective functions \cite{Ganguli2014, Wang2016NIPS} when Fisher information was assumed to be constant.
Here we see that, for any neural code that satisfies our general assumptions, under any objective function, when the code is adapted to a uniform stimulus distribution, it will deviate from flat-Fisher equalization if and only if there are constraints present on the neural activity.
This property allows us to test whether such constraints are present: by allowing the neural code to adapt to $p_s(s) \propto 1$ and measuring $\smash{ p_{\hat{\theta}}(\hat{\theta}) }$ we can test whether the code is constrained ($p_{\hat{\theta}}(\hat{\theta}) \not \propto 1$ ) or unconstrained ($p_{\hat{\theta}}(\hat{\theta}) \propto 1$). 

This test can also be phrased in terms of the Fisher information in the code about the stimulus, $I_s(s)$.
Eqs.~(\ref{eq:logIs}) and (\ref{eq:uniformIs}) reveal that, when unconstrained, $I_s(s) \propto p_s(s)^2$, so that there is more information about more likely stimuli.
This is intuitive because the histogram equalization solution spans a wider dynamic range of parameters across more likely stimuli.
In fact, $p_{\hat \theta}(\hat \theta) \propto p_s(s) / \sqrt{I_s(s)}$  (Eq.~(\ref{eq:flatSpace})) shows that in this case, the flat-Fisher space parameter density measures the local deviation from histogram equalization.

\subsubsection{Non-log objective functions result in stimulus distribution dependence} 
\label{sec:warping} 
Our second test focuses on the \emph{adaptation} of the neural code, which is to say, the differences in the neural code when it is optimized for different stimulus distributions, $p_s(s)$.
As we saw in the unconstrained solution (Sec.~\ref{sec:unconstrained} and Fig.~\ref{fig:analyticResults}b), a major impact of the objective function is that it changes how the optimal solution adapts to different stimulus distributions, with each objective function having its own characteristic adaptation behavior.
The log objective function is unique from an adaptation perspective because its optimal parameter distribution $\smash{ p_{\hat \theta}(\hat \theta) }$ is invariant to changes in the stimulus distribution.
This happens because the optimal stimulus-to-parameter mapping $\smash{\hat \theta(s)}$ adapts to different stimulus distributions in a way that compensates for these differences, leaving $p_{\hat \theta}$ unchanged (see Fig.~\ref{fig:analyticResults}a).
In this unconstrained case, the log objective function will always achieve flat-Fisher histogram equalization, while non-log objective functions will not be histogram equalizing except for uniform stimulus distributions.
This can be seen by expressing the solution as:
\begin{equation}
    f' \left( \frac{p_s(s)}{p_{\hat \theta}( \hat \theta(s))} \right) p_s(s) \propto 1.
\end{equation}
This shows that the $p_s(s)$ dependency only cancels if $f'(x) \propto 1/x$, which requires a logarithmic objective function.
As we show in the supplementary material (Sec. 4.1), this property extends to general constraints: the logarithm is the only objective function whose parameter distributions are invariant to changes in the stimulus distribution, regardless of constraints.
This is essentially because the stimulus distribution still enters these (differential) equations through a term of the same form as above, $f' \left( p_s(s) / p_{\hat \theta}( \hat \theta) \right) p_s(s)$. 
Thus, neural code adaptation allows us to test whether or not the objective function is log: regardless of the presence or absence of constraints on neural activity, if changing the stimulus distribution does not impact the parameter distribution $p_{\hat \theta}(\hat \theta)$ of the optimized code, then the objective function is $\smash{f(\sqrt{I}) \propto \log I}$; otherwise it is not.

The log objective function is important because, under the right conditions, it approximates the mutual information \cite{Brunel1998, Wei2016}, and so should be a reasonable objective for coding in early sensory areas. 
In fact, when adapted to Gaussian distributions of visual angular velocity stimuli with different variances, the stimulus-to-neural firing rate relationship of H1 neurons in the blowfly visual system only differs in the scaling the of stimulus \cite{Brenner2000, Fairhall2000, Fairhall2001}.
In other words, the activity distributions are preserved.
By our log objective function test, this is evidence for a logarithmic (i.e. mutual information maximizing) objective function.

\subsubsection{A closed-loop experimental paradigm to identify both objective and constraint functions}

The tests derived thus far support detecting the presence of constraints and the presence of log-objective functions, but do not allow us to fully characterize the objective or constraint function that a code is optimized for.
These previous tests suggest that the constraint impacts deviations from histogram equalization, while the objective function impacts adaptation of the neural code to different stimulus distributions.
In order to disentangle their impacts on the optimized code, we first 'remove' the effects of adaptation by finding a fixed point of the neural adaptation. At this fixed point, which we will introduce in detail below, the optimal coding solution becomes independent of the objective function, and thus allows us to measure the constraint function in isolation.
Once this is achieved, we can measure adaptation to another non-fixed point stimulus distribution, which depends on both constraint and objective function, and allows us to determine the latter.

Let us first describe what we mean by a \emph{fixed point} of the neural code adaptation.
Because we are examining how the neural code adapts to different stimulus distributions, we can think of this adaptation as a mapping from distributions of stimuli to distributions of neural activity parameters: ${p_s(s) \to p_{\hat \theta}( \hat \theta(s) )}$.
A fixed point of such a mapping is a stimulus distribution, $p_s(s)$, such that after adaptation to $p_s(s)$, the neural parameter distribution is proportional to this same distribution: $\smash{ p_{\hat{\theta}}(\hat{\theta}(s)) \propto p_s (s) }$.
This fixed point can equivalently be characterized by $\smash{\hat \theta(s) = a s + b}$: a change in stimulus induces the same change in activity parameters regardless of the starting stimulus, or via Eq.~(\ref{eq:flatSpace}) by $I_s(s) \propto 1$: at the fixed point, the neural code contains the same Fisher information about all stimuli.
We can find the fixed point in closed form (see also supplementary material) using the proportionality of stimulus and parameter probability densities.
Note that in  Eq.~(\ref{eq:usedObj}), the argument to the object function equals the ratio of probability densities $\smash{ p_s(s) / p_{\hat \theta}( \hat \theta(s))}$, which is constant at the fixed point.
This property extends to the solutions of the optimization, rendering the fixed point solution independent of the objective function.
The fixed point density is given by
\begin{equation}
    p_{\hat{\theta}}(\hat{\theta}(s)) \propto p_s (s) \propto \exp \left( -\lambda C_{\hat \theta} \left( \hat{\theta}(s) \right) \right),
\end{equation}
where $C_{\hat \theta}(\hat \theta) = C( \theta( \hat \theta) )$ is the constraint function expressed as a function of the flat-Fisher parameter.
At this stimulus distribution, the distribution of neural activity parameters is equal to the stimulus distribution and both are equivalent to the log-objective function solution (Eq.~(\ref{eq:logpdf}) expressed in flat-Fisher parameters).
This intuitively makes sense because the log-objective function solution is independent of the stimulus distribution, such that it must always occupy the fixed point.

Crucially, the fixed point solution is the same regardless of the objective function that the code is optimizing.
Thus, finding the fixed point gives us an objective-independent characterization of the constraint, $\smash{ C_{ \hat \theta}(\hat \theta)}$.
With the constraint function in hand, we can then probe the neural code at another stimulus distribution to find the objective function.
For example, with a uniform stimulus distribution, Eq.~(\ref{eq:uniformpdf}), if we know  $C_{\hat \theta}(\cdot)$ and $p_{\hat \theta}(\cdot)$, then $\smash{\hat f^{-1}(\cdot)}$ can be found by regression.

The only remaining piece in the puzzle is to find the fixed point stimulus distribution.
This can be accomplished with a \textit{closed-loop experimental paradigm}, which is in essence a fixed point iteration: the neural activity measured in response to stimuli drawn from one stimulus distribution determines the stimulus distribution that will be presented next.
Specifically, we allow the neural code to adapt to an initially uniform stimulus distribution, $p_s(s) \propto 1$, and then measure neural responses $\bm r_i$ to stimuli $s_i \sim p_s$.
From these responses, we fit the flat-Fisher space distribution, $p_{\hat \theta}$, which then becomes the new stimulus distribution. 
The process of re-adaptation, measurement, and updating the stimulus distribution repeats until there are no changes to the stimulus distribution.
At this point, $p_{\hat \theta} \propto p_s$, and the adaptation fixed point is achieved.
The constraint can then be fit from the fixed point at the last step of this iteration and, given this constraint, the objective can be fit to the results from the first step, which had a uniform stimulus distribution.  
There is one edge case: if the code is unconstrained, then we know from the constraint test that the flat-Fisher parameter distribution will be uniform when the stimulus distribution is uniform. As a consequence, the iteration will terminate on the first step.
If this is the case, another stimulus distribution is needed to identify the objective function.
Because we now know that the code is unconstrained, this can be accomplished using any stimulus distribution and the general unconstrained solution, Eq.~(\ref{eq:C0pdf}).

\subsection{Estimating flat-Fisher space parameter distributions from neural activity data}
\label{sec:fitting}

\begin{figure}[tbh]
    \centering
    \caption{Fitting (a) $\theta(s)$ and (b) flat space distribution, $p_{\hat \theta}(\hat \theta(s))$ to simulated activity data (black dashes). Shown are 10 fits to the optimal solution with uniform stimulus distribution, Poisson activity, $\theta \in [5,55]$, $f(x)=-\sqrt{x}$,  $C(\theta)=\sin( 2\pi / 50 \; \theta ) + 1$ and $M = 0.75$ from 2000 trials each.} \label{fig:fitting}
    \includegraphics{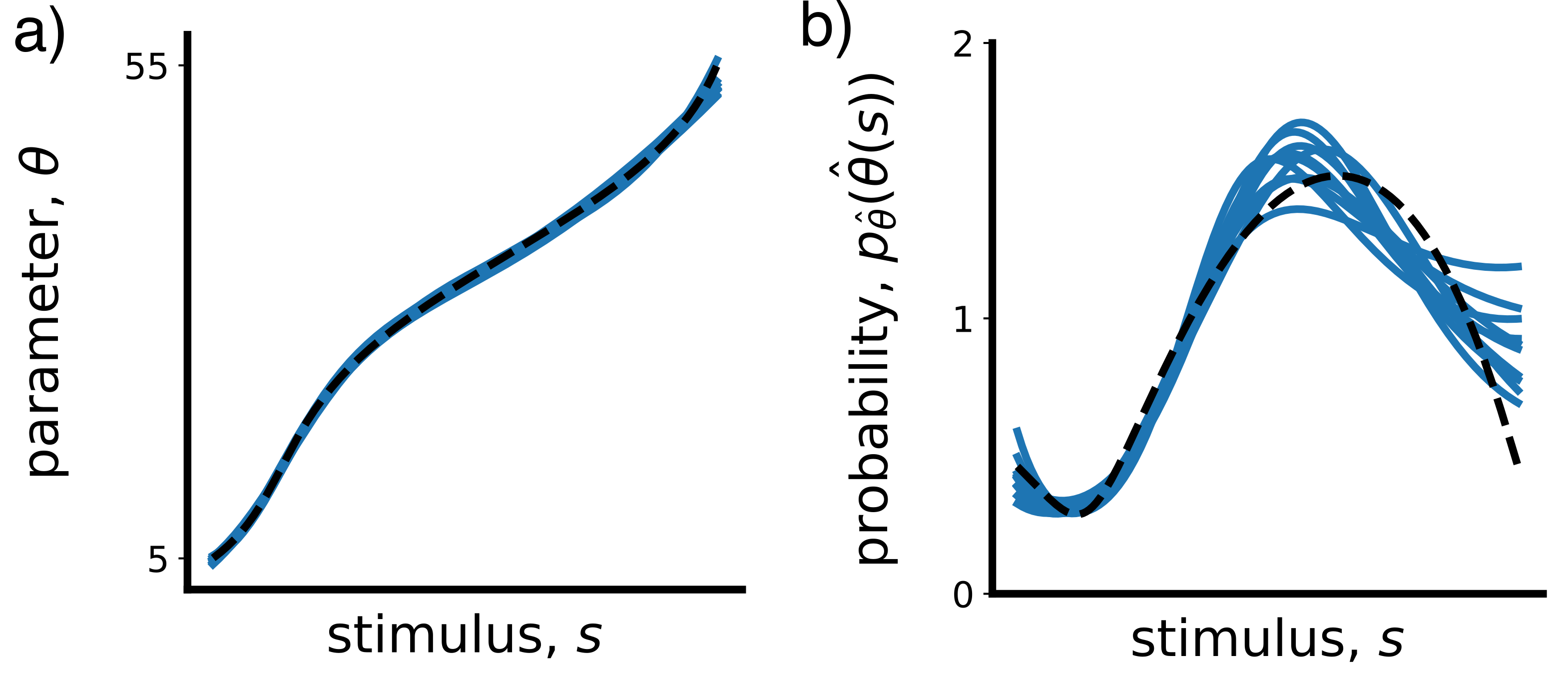}
\end{figure}

Each of the above tests requires access to the distribution of  flat-Fisher space neural activity parameters $\smash{\hat{\theta}(s)}$, sampled across stimuli.
As we show here, these activity parameter distributions can in principle be recovered from neural activity data, which consists of a series of stimulus, response pairs: $\left\{ s_i, \bm r_i \right\}_i$.
For this purpose we assume that the noise in the neural encoding distribution $p(\bm r|s)$ is known, or is well-described by an exponential family distribution, $p(\bm{r} | s) = \exp ( \bm{\eta}(s)^T \bm{T}(\bm{r}) - A(\bm{\eta}(s)))$, where the sufficient statistics $\bm{T}(\bm{r})$ are assumed to be fixed and known, and $A \left( \bm{\eta}(s) \right) = \int \exp \left( \bm{\eta}(s)^T \bm{T} (\bm{r}) \right) \ud \bm{r}$ is the log-normalizer.
We will require gradients and Hessians of the log normalizer, which we will take to be known for our model family, or computed by sampling the moments of the sufficient statistics at fixed $\bm \eta$ values.
Given such a model of the neural activity, our task is to fit the neural activity parameter submanifold $\bm \eta(s)$, which we do by GLM regression, modeling $\bm{\eta}(s)$ by a neural network and estimating the network's weight parameters $\bm{w}$ by ascending the log-likelihood gradient, given by $\frac{\partial L}{\partial w} = \left( \sum_i \bm{T}(\bm r_i) - \frac{\partial A(\eta(s_i))}{\partial \bm \eta} \right) \frac{\partial \bm \eta}{\partial w}$.

Once $\bm \eta(s)$ is known, we can use it to find the flat-Fisher space parameter distribution $p_{\hat{\theta}} ( \hat{\theta} )$.
We can do so from Eq.~(\ref{eq:flatSpace}), which shows that it is sufficient to know the stimulus Fisher information and stimulus distribution to recover the flat-Fisher space parameter distribution, $\smash{p_{\hat \theta}(\hat \theta(s)) }$.
As a proxy for stimulus Fisher information in neural activity, we use the stimulus Fisher information of the fitted encoding model, which can be estimated by forward differentiation of the model to obtain $d\bm \eta / ds$. 
By transforming inputs through the stimulus CDF prior to training, $\tilde s = \textrm{CDF}_s(s)$, the flat space distribution can be obtained directly from these Jacobians (again via Eq.~(\ref{eq:flatSpace})) by ${p_{\hat \theta}(\hat \theta( \tilde s)) \propto 1 / ( \sqrt{ (d\bm \eta / d \tilde s)^T I_{\bm \eta} ( d\bm \eta / d \tilde s) } ) }$.
The parameter Fisher information in our model, $I_{\bm \eta}$ is given by the Hessian of the log normalizer.
In Fig.~\ref{fig:fitting} we illustrate feasibility of this approach in a simple example.

\section{Discussion}
The match between efficient coding predictions and behavior of adaptive codes is no accident.
By requiring first-order stability, efficient coding captures feature of codes that have achieved a steady-state of adaptation.
Vice verse, by probing the adaptation behaviors of neural codes, we can characterize these codes in terms of what they are optimizing for.
Here we characterized a relatively broad class of efficient coding problems, but there is still much to be achieved.
For example the assumptions of one-dimensional stimuli and parameter are limiting and unrealistic.
Optimizing multiple parameters would allow the shape of the parameter subspace to change, rather than simply distance, raising the question whether shape or scale optimizations dominate in different forms of adaptation.
Additionally, the adaptation is not always at steady-state. In natural scenes there are many different, simultaneously active timescales that the neural code may be adapting to at any given time.

\section*{Statement of Broader Impact}
This work builds toward improved understanding of the neural code, and thus, a better understanding of operation of the nervous systems and the behavior of humans and other animals. 
Such understanding is important for neural prostheses and may lead to novel treatment options of human brain diseases.
One potential risk is that a better understanding of neural coding, particularly value coding, lends itself to abuse for human behavioral modification. This is not an immediate concern for the work presented here.

\section*{Acknowledgments and Disclosure of Funding}
We would like to thank Johannes Bill and Haim Sompolinsky for fruitful discussions.

This work was supported by funding from the National Institutes of Health (R01MH115554, J.D.), a Scholar Award in Understanding Human Cognition by the James S. McDonnell Foundation (grant\# 220020462, J.D.), and a Lefler Small Grant from the Edward R. and Anne G. Lefler Center at Harvard Medical School (L.R.).

\bibliographystyle{unsrt}
\bibliography{combinedDoc.bib}

\pagebreak


\section*{Supplementary material}

# Supplementary Material: Adaptation Properties Allow Identification of Optimized Neural Codes

Luke Rast, Jan Drugowitsch

## 1 General Approach

As discussed in the main text, we assume a neural encoding model $p(\boldsymbol{r}|s)$, which specifies how neural (population) activity $\boldsymbol{r}$ depends on a one-dimensional stimulus $s$, and is parametrized by a scalar-valued, deterministic function $\theta(s)$ such that

$$p(\boldsymbol{r}|s) = p(\boldsymbol{r}|\theta(s)), \tag{1}$$

where $\theta(s)$ is increasing in $s$. We assume that the code is efficient in that it maximizes the average across $s$ of a function $f(\cdot)$ of the square root of the Fisher information in $\boldsymbol{r}$ about $s$, $I_s(s) = \mathbb{E}_{p(\boldsymbol{r}|s)}\left[\left(\frac{\mathrm{d}}{\mathrm{d}s}\log p(\boldsymbol{r}|s)\right)^2\right]$, while satisfying, on average, some constraint $h(\boldsymbol{r}) \leq M$, that is

$$\max \; \mathbb{E}_{p_s(s)}\left[f\left(\sqrt{I_s(s)}\right)\right], \qquad \text{s.t.} \quad \mathbb{E}_{p(\boldsymbol{r}|s)p_s(s)}[h(\boldsymbol{r})] \leq M. \tag{2}$$

Specific assumptions about $f(\cdot)$ and $h(\cdot)$ are discussed in the main text, and will be revisited further below.

### 1.1 Deriving the efficient coding objective in terms of $\theta(s)$

Here we derive the $\theta(s)$ optimization from our assumptions. Because $\theta$ is a deterministic remapping of $s$, an application of the chain rule shows that the Fisher information in $\boldsymbol{r}$ about $s$ is related to the Fisher information in $\boldsymbol{r}$ about $\theta$ by

$$I_s(s) = I_\theta\big(\theta(s)\big)\left(\frac{\mathrm{d}\theta}{\mathrm{d}s}\right)^2. \tag{3}$$

Note that $I_\theta(\theta) = \mathbb{E}_{p(\boldsymbol{r}|\theta)}\left[\left(\frac{\mathrm{d}}{\mathrm{d}\theta}\log p(\boldsymbol{r}|\theta)\right)^2\right]$ is determined only by features of the encoding model. Turning to the constraint,

$$\mathbb{E}_{p_s(s)p(\boldsymbol{r}|s)}[h(\boldsymbol{r})] = \mathbb{E}_{p_s(s)}\left[\int h(\boldsymbol{r})p(\boldsymbol{r}|\theta(s))\mathrm{d}\boldsymbol{r}\right] := \mathbb{E}_{p_s(s)}\left[C(\theta(s))\right], \tag{4}$$

where we have defined $C(\theta) := \mathbb{E}_{p(\boldsymbol{r}|\theta)}[h(\boldsymbol{r})]$. Finally, we made two additional coding assumptions on $\theta(s)$, namely that $\theta(s)$ is increasing

$$\frac{\mathrm{d}\theta}{\mathrm{d}s} > 0 \tag{5}$$

and that the $\theta(s)$ curve has finite extent $L_\theta$ in $\theta$ space

$$\int \frac{\mathrm{d}\theta}{\mathrm{d}s}\mathrm{d}s = \int \mathrm{d}\theta = L_\theta. \tag{6}$$

Taking the assumed objective function, Eq. (2), and (i) plugging Eq. (3) into the objective, (ii) plugging Eq. (4) into the constraint, (iii) introducing the encoding specific constraints, Eqs. (5) and (6), yields the primary optimization problem from the main text:

$$\max_{\theta(s)} \; \mathbb{E}_{p_s(s)}\left[f\left(\sqrt{I_\theta(\theta)}\frac{\mathrm{d}\theta}{\mathrm{d}s}\right)\right] \quad \text{s.t.} \quad \mathbb{E}_{p_s(s)}[C(\theta(s))] \leq M, \quad \frac{\mathrm{d}\theta}{\mathrm{d}s} > 0, \quad \int \frac{\mathrm{d}\theta}{\mathrm{d}s}\mathrm{d}s = L_\theta. \tag{7}$$

## 1.2 Assumptions about functional forms

The features of our efficient coding problem, $p_s(s)$, $f(\sqrt{I})$, $I_\theta(\theta)$, and $C(\theta)$, are allowed to be quite flexible, but are still subject to some constraints, as listed here:

- $p_s(s)$ is assumed to be continuous, and to have no zero-probability 'gaps'. This latter assumption is for convenience, and can be removed without changing the results.

- $f(\sqrt{I})$ is assumed to be continuous and monotonically increasing. Given these assumptions, we shown in the next section that $f(\cdot)$ must additionally be concave and asymptotically sublinear for the optimization-problem to be well-behaved.

- $I_\theta(\theta)$ is assumed to be continuous and non-zero (note that Fisher information is always $\geq 0$).

- $C(\theta)$ is assumed to be continuous and non-negative. Note that non-negativity follows from non-negativity of $h(\boldsymbol{r})$.

## 1.3 Concavity and sub-linearity of the objective function

The requirement that the objective function is concave and asymptotically sub-linear can be demonstrated most directly in an optimal control approach to solving the variational optimization. Here, we interpret the efficient coding problem as optimizing a control $u(s)$ on how fast $\theta$ is increasing at each stimulus, that is, $u(s) := \frac{d\theta}{ds}$. Note that in this formulation, the stimulus $s$ plays the role of time in a traditional optimal control setting. The last two constraints in Equation 7 act as a restriction on the control set, and a boundary condition for state $\theta$, respectively. For such optimal control problems, the Pontryagin maximality principle (see [1, 3]) states that, if we form the Hamiltonian

$$H(s, \theta(s), q(s), u(s)) = q(s)u(s) + f\left(u(s)\sqrt{I_\theta(\theta(s))}\right)p_s(s) + \lambda C(\theta(s))p_s(s), \quad (8)$$

by adding to the objective function a term composed of the product between the dynamics ($d\theta/ds = u$) and a 'momentum' co-state, $q(s)$, then the optimal trajectory will be given by a local optimization of $u(s)$,

$$u^*(s) = \mathrm{argmax}_{u(s)} H(s, \theta(s), q(s), u(s)), \quad (9)$$

in conjunction with Hamilton's equations,

$$\frac{d\theta}{ds} = \frac{\partial H}{\partial q}, \quad \frac{dq}{ds} = -\frac{\partial H}{\partial \theta}. \quad (10)$$

Here we focus on the control maximization:

$$\max_{u(s)} \quad q(s)u(s) + f\left(u(s)\sqrt{I_\theta(\theta)}\right)p_s(s). \quad (11)$$

Recall that $u(s) = d\theta/ds$ is strictly positive, and $f(\cdot)$ is increasing. This means that, in order to have a maximum where $u(s) \neq \infty$, $q(s)$ must be negative everywhere, and $f(\cdot)$ must increase sublinearly as $u(s) \to \infty$. Furthermore, because $u(s) \in (0, \infty)$ cannot be at the boundaries, we have a second order condition for maximality:

$$\frac{\partial^2}{\partial u^2}\left(q(s)u(s) + f\left(u(s)\sqrt{I_\theta(\theta)}\right)p_s(s)\right) = f''\left(u(s)\sqrt{I_\theta(\theta)}\right)I_\theta(\theta)p_s(s) < 0. \quad (12)$$

The non-negativity of $I_\theta(\theta)$ and $p_s(s)$ implies that maxima are only possible in places where $f(\cdot)$ is concave. This restriction does not fully rule out non-concave objectives, for which optimal solutions could transition discontinuously between concave regions of the objective function. This behavior might be interesting, but we limit the analysis to continuous $u(s)$. That is to say, we require our solution to be well-behaved, in the sense that $\theta(s)$ is $C^1$, continuous and with continuous derivatives. This is the standard requirement for optimization by Euler-Lagrange. Hence, in order for solutions $\theta(s)$ to be $C^1$, the objective function must be concave and asymptotically sublinear as its argument increases.

# 2 Finding efficient coding solutions

Here we derive the solution to the efficient coding problem above. We first outline the general approach, leading to a system of ordinary differential equations that describe the efficient coding solution. From this we derive in turn the three special-case analytic solution described in the main text.

## 2.1 General optimal solution

There are several ways to solve the general optimization problem. Here we walk through an approach that is particularly intuitive, but requires some setup. Start with the general optimization problem (Eq. (7)),

$$\max_{\theta(s)} \mathbb{E}_{p_s(s)} \left[ f\left( \sqrt{I_\theta(\theta)} \frac{d\theta}{ds} \right) \right] \quad \text{s.t.} \quad \mathbb{E}_{p_s(s)}[C(\theta)] \leq M, \quad \frac{d\theta}{ds} > 0, \quad \int \frac{d\theta}{ds} ds = L_\theta. \quad (13)$$

Because $\theta(s)$ is continuous and strictly increasing, we know that $\theta$ is an invertible function of $s$, and can therefore be re-expressed in terms of cumulative distribution functions or probability density functions

$$\theta(s) = \text{CDF}_\theta^{-1}(\text{CDF}_s(s)); \qquad p_\theta(\theta)\mathrm{d}\theta = p_s(s)\mathrm{d}s. \quad (14)$$

Equivalently,

$$s(\theta) = \text{CDF}_s^{-1}(\text{CDF}_\theta(\theta)); \qquad \frac{\mathrm{d}\theta}{\mathrm{d}s} = \frac{p_s(s(\theta))}{p_\theta(\theta)}. \quad (15)$$

The idea here is to transform the optimization of the function $\theta(s)$ into an optimization of $p_\theta(\theta)$, the probability density function. To do this, we first change the variable of integration from $p_s(s)\mathrm{d}s$ to $p_\theta(\theta)\mathrm{d}\theta$ and substitute the ratio of probability density functions in place of the $\theta$ derivative,

$$\max_{p_\theta(\theta)} \int_0^{L_\theta} f\left( \frac{p_s(s(\theta))}{p_\theta(\theta)} \sqrt{I_\theta(\theta)} \right) p_\theta(\theta)\mathrm{d}\theta \quad \text{s.t.} \quad \int_0^{L_\theta} C(\theta)p_\theta(\theta)\mathrm{d}\theta \leq M, \quad \int_0^{L_\theta} p_\theta(\theta)\mathrm{d}\theta = 1 \quad (16)$$

In this formulation, the final two (parameter) constraints have been absorbed into assumptions on the probability density function. First, $p_\theta(\theta)$ is positive. Second, the length of $\theta$-space is $L_\theta$, and $p_\theta$ is normalized over this length. Here we have WLOG set the lower bound of $\theta$ to zero. Note that $s(\theta) = \text{CDF}_s^{-1}(\text{CDF}_\theta(\theta))$ has its own $\theta$ dependence, which introduces some notational complexity. To handle this complexity, we define the additional functions $\alpha(\theta)$ and $\pi_s(\alpha)$, given by

1. $\text{CDF}_\theta(\theta) = \alpha(\theta)$,
2. $\frac{d\alpha}{d\theta} = p_\theta(\theta)$,
3. $p_s(s(\theta)) = p_s(\text{CDF}_s^{-1}(\text{CDF}_\theta(\theta))) = \pi_s(\alpha(\theta))$.

This turns the optimization problem into

$$\max_{\alpha(\theta), p_\theta(\theta)} \int_0^{L_\theta} f\left( \frac{\sqrt{I_\theta(\theta)} \pi_s(\alpha(\theta))}{p_\theta(\theta)} \right) p_\theta(\theta)\mathrm{d}\theta \quad \text{s.t.} \quad \begin{cases} \int C(\theta)p_\theta(\theta)\mathrm{d}\theta \leq M, \\ \int p_\theta(\theta)\mathrm{d}\theta = 1, \\ \frac{d\alpha}{d\theta} = p_\theta(\theta) \end{cases} \quad (17)$$

Here we are optimizing both of the functions $\alpha(\theta)$ and $p_\theta(\theta)$ under the constraint that $\frac{d\alpha}{d\theta} = p_\theta(\theta)$ (by the definition of $\alpha(\theta)$), that is, that they specify the cumulative and probability density functions of the same distribution. The $p_\theta(\theta)$ optimization has a familiar form. Along with the top two constraints, it takes the from of a constrained minimization of the $-f$ divergence [2] between $p_\theta$ and $\sqrt{I_\theta(\theta)}\pi_s(\alpha(\theta))$. In this problem, however, the target distribution is also dependent on the distribution that we are optimizing. We will see that this target adaptation is what accounts for stimulus-distribution dependent adaptation.

Optimizing both $\alpha(\theta)$ and $p_\theta(\theta)$, the Euler-Lagrange equations result in the system of ordinary differential equations,

$$\hat{f}\left(\frac{\sqrt{I_\theta(\theta)}\pi_s(\alpha(\theta))}{p_\theta(\theta)}\right) = \lambda C(\theta) + Z - \gamma(\theta) \tag{18}$$

$$\frac{d\gamma}{d\theta} = f'\left(\frac{\sqrt{I_\theta(\theta)}\pi_s(\alpha(\theta))}{p_\theta(\theta)}\right)\sqrt{I_\theta(\theta)}\pi'_s(\alpha(\theta)), \tag{19}$$

$$\frac{d\alpha}{d\theta} = p_\theta(\theta). \tag{20}$$

Here, $\lambda$, $Z$, and $\gamma(\theta)$ are Lagrange multipliers in charge of enforcing, respectively, the constraint on $C(\theta)$, normalization of $p_\theta(\theta)$, and the equality between $p_\theta$ and the derivative of the $\theta$ CDF, $\alpha(\theta)$. The constraint on $C(\theta)$, and the required normalization, can be satisfied by solving (numerically if necessary) for the values of $\lambda$ and $Z$. The objective function appears here in two forms: the derivative, $f'(\cdot)$, in Eq. (19), and the Legendre transform, $\hat{f}(\cdot)$, in Eq. (18), which is defined as $\hat{f}(x) = f'(x)x - f(x)$.

## 2.2 Special-case analytic solutions

We now derive each of the special cases in turn.

### 2.2.1 Log-objective function, $f(\sqrt{I}) \propto \log I$

Setting $f(x) = \log x$, implies that $f'(x) = 1/x$, and $\hat{f}(x) = 1 - \log x$, in which case the $\gamma(\theta)$ dynamics, Eq. (19), become

$$\frac{d\gamma}{d\theta} = \frac{\pi'_s(\alpha(\theta))}{\pi_s(\alpha(\theta))}p_\theta(\theta) = \frac{d}{d\theta}\log \pi_s(\alpha(\theta)) \tag{21}$$

Hence, $\gamma(\theta) = \log \pi_s(\alpha(\theta)) + c_0$ with constant $c_0$. Plugging this result into Eq. (18) cancels the $\pi_s$ dependence on the right-hand side and, absorbing constants into $Z$, the remaining equation can be solved to yield

$$p_\theta(\theta) = Z^{-1}\sqrt{I_\theta(\theta)}\exp(-\lambda C(\theta)). \tag{22}$$

### 2.2.2 Unconstrained optimization, $M \to \infty$

For convenience denote $G(\theta) = \frac{\sqrt{I_\theta(\theta)}\pi_s(\alpha(\theta))}{p_\theta(\theta)}$. In Eq. (18), setting $C(\theta) = 0$ (or $\lambda = 0$ if eg. a constraint is present, but not binding) means that $\gamma(\theta) = Z - \hat{f}(G(\theta))$. Furthermore, note that from the definition $\hat{f}(x) = f'(x)x - f(x)$,

$$\frac{d\gamma}{d\theta} = -\frac{d}{d\theta}\hat{f}(G(\theta)) = -f''(G(\theta))G(\theta)G'(\theta). \tag{23}$$

This $\gamma$ derivative can be plugged directly into Eq. (19), yielding:

$$-\frac{f''(G(\theta))}{f'(G(\theta))}G'(\theta) = \frac{\pi'_s(\alpha)}{\pi_s(\alpha)}p_\theta(\theta) \tag{24}$$

Equivalently,

$$\frac{d}{d\theta}\log f'(G(\theta)) = \frac{d}{d\theta}\log \frac{1}{\pi_s(\alpha(\theta))}. \tag{25}$$

Hence,

$$f'\left(\frac{\sqrt{I_\theta(\theta)}\pi_s(\alpha(\theta))}{p_\theta(\theta)}\right) \propto \frac{1}{\pi_s(\alpha(\theta))}, \tag{26}$$

from which it follows that

$$p_\theta(\theta) = \sqrt{I_\theta(\theta)}\frac{p_s(s(\theta))}{f'^{-1}\left(\frac{Z}{p_s(s(\theta))}\right)}, \tag{27}$$

the form shown in the main text.

4

### 2.2.3 Uniform stimulus probability density, $p_s(s) \propto 1$

From the discussion above, when the stimulus distribution is uniform, that is $p_s(s) = p_s$, and $\pi_s(\cdot) \propto 1$, the $-f$ divergence interpretation becomes clear. In this case, $\pi'_s(\alpha(\theta)) = 0$, making $\gamma(\theta)$ a constant that can be absorbed into $Z$. Then, solving Eq. (18) yields

$$p_\theta(\theta) = \sqrt{I_\theta(\theta)} \frac{p_s}{\hat{f}^{-1}(\lambda C(\theta) + Z)}. \tag{28}$$

### 2.3 Multiple constraints

In the above derivation, the constraint $C(\theta)$ passes through unchanged to give a factor of $\lambda C(\theta)$ in Eq. (18). The same will occur with multiple constraints, to give a term $\boldsymbol{\lambda}^T \boldsymbol{C}(\theta)$ in place of $\lambda C(\theta)$ in all of the above solutions. Satisfying the constraints then requires solving a higher dimensional system of equations for the values of the Lagrange multipliers, but the results remain qualitatively unchanged.

## 3 Flat-Fisher space reparameterization

As discussed in the main text, transformation to the flat-Fisher space facilitates an understanding of the efficient coding solutions that is invariant to our parameterization of the neural activity. Here we derive this remapping, apply it to special-case solutions and the overall problem formulation, and then derive the relationship between the flat-Fisher space parameter distribution and the Fisher information in stimulus space.

### 3.1 Defining the reparametrization

We want our reparameterization to satisfy

$$d\hat{\theta} = \sqrt{I_\theta(\theta)} d\theta. \tag{29}$$

First we have to ensure that such a reparameterization is well behaved and well defined. Because $\sqrt{I_\theta(\theta)}$ is strictly positive ("strictly" by assumption), we can always solve this differential equation to find a strictly monotonic (increasing) function $F$:

$$\hat{\theta}(\theta) = F(\theta) = \int_{\theta_0}^{\theta} \sqrt{I_\theta(\theta')} d\theta'. \tag{30}$$

This reveals that such a reparameterization will be invertible and thus well behaved. However, the transformation is only defined up to the integration constant. In fact, the Fisher information can be made constant by any transformation of the form $d\hat{\theta} = a\sqrt{I_\theta(\theta)}d\theta$ with constant $a$:

$$I_{\hat{\theta}}\left(\hat{\theta}\right) = I_\theta\left(\theta\left(\hat{\theta}\right)\right) \left(\frac{d\theta}{d\hat{\theta}}\right)^2 = \frac{1}{a^2} \tag{31}$$

Therefore, the flat-Fisher space is well-defined only up to shifting the origin and re-scaling the length. For our purposes, neither of these will matter: the scale only serves to change the normalization of the distribution $p_{\hat{\theta}}$, and we are free to choose the origin of $\hat{\theta}$.

### 3.2 Reparameterization of special-case solutions

As noted in the main text, all of the special-case solutions are scaled by a factor of $\sqrt{I_\theta}$. Owing to the invertible mapping between $\theta$ and $\hat{\theta}$, this scaling can be removed by reparameterization to the flat-Fisher space,

$$p_{\hat{\theta}}\left(\hat{\theta}\right) = p_\theta\left(\theta\left(\hat{\theta}\right)\right) \frac{d\theta}{d\hat{\theta}} = \frac{p_\theta\left(\theta\left(\hat{\theta}\right)\right)}{\sqrt{I_\theta\left(\theta\left(\hat{\theta}\right)\right)}}. \tag{32}$$

This removes $I_\theta$ terms from the solution, but does require reparameterizing the constraint to $C\left(\theta\left(\hat{\theta}\right)\right) = C\left(F^{-1}\left(\hat{\theta}\right)\right) := C_{\hat{\theta}}\left(\hat{\theta}\right)$.

### 3.3 Reparameterization of problem formulation

Explicit $\sqrt{I_\theta}$ dependence can be fully removed from the optimization problem by reparameterizing prior to solving. We first note that

$$\frac{\mathrm{d}\hat{\theta}}{\mathrm{d}s} = \frac{\mathrm{d}\hat{\theta}}{\mathrm{d}\theta}\frac{\mathrm{d}\theta}{\mathrm{d}s} = \sqrt{I_\theta}\frac{\mathrm{d}\theta}{\mathrm{d}s}. \tag{33}$$

and that

$$\frac{\mathrm{d}\hat{\theta}}{\mathrm{d}s} = \frac{p_s(s)}{p_{\hat{\theta}}(\hat{\theta})} = \frac{\sqrt{I_\theta(\theta)}p_s(s)}{p_\theta(\theta)}. \tag{34}$$

From Eq. (33) and the fact that $\sqrt{I_\theta}$ is strictly positive, it follows that, $\mathrm{d}\theta/\mathrm{d}s > 0 \iff \mathrm{d}\hat{\theta}/\mathrm{d}s > 0$. Thus, the positivity constraint on the $\theta$ derivatives can be replaced by a positivity constraint on $\hat{\theta}$ derivatives, and, plugging Eq. (33) in the objective function in the optimization, Eq. (7), yields

$$\max_{\hat{\theta}(s)} \mathbb{E}_{p_s(s)}\left[f\left(\frac{\mathrm{d}\hat{\theta}}{\mathrm{d}s}\right)\right] \quad \text{s.t.} \quad \mathbb{E}_{p_s(s)}\left[C_{\hat{\theta}}\left(\hat{\theta}\right)\right] \leq M, \quad \frac{\mathrm{d}\hat{\theta}}{\mathrm{d}s} > 0, \quad \int \frac{\mathrm{d}\theta}{\mathrm{d}s}\mathrm{d}s = L_\theta. \tag{35}$$

Equation 34 ensures that the solution approach described above goes through unchanged, replacing $\sqrt{I_\theta(\theta)}/p_\theta(\theta)$ by $1/p_{\hat{\theta}}(\hat{\theta})$.

One step remains for transformation to the flat-Fisher space: satisfaction of the length constraint on $\theta$. The idea here is to make use of the invertible mapping $\theta \leftrightarrow \hat{\theta}$, such that if we move from a starting location in $\theta$-space, $\theta_0$, by some distance $L_\theta$, then, in doing so, we will have moved from the corresponding location in $\hat{\theta}$-space, $\hat{\theta}_0$, by a distance $L_{\hat{\theta}}$. Regardless of the value of $L_\theta$, the invertible mapping provides the transformation from $\theta$-distances to $\hat{\theta}$-distances. More specifically,

$$L_\theta = \int_{s_0}^{s_1} \frac{\mathrm{d}\theta}{\mathrm{d}s}\mathrm{d}s = \int_{\theta_0}^{\theta_1} \mathrm{d}\theta = \theta_1 - \theta_0, \tag{36}$$

while

$$L_{\hat{\theta}} = \int_{s_0}^{s_1} \frac{\mathrm{d}\hat{\theta}}{\mathrm{d}s}\mathrm{d}s = \int_{\theta_0}^{\theta_1} \sqrt{I_\theta(\theta)}\mathrm{d}\theta = F(\theta_1) - F(\theta_0). \tag{37}$$

So, given a starting point, $\theta_0$ and a $\theta$-length, $L_\theta$, we can find the corresponding $\hat{\theta}$-length by:

$$L_{\hat{\theta}} = F(\theta_0 + L_\theta) - F(\theta_0). \tag{38}$$

The same logic provides a transformation from $\hat{\theta}$-lengths to $\theta$-lengths:

$$L_\theta = F^{-1}(\hat{\theta}_0 + L_{\hat{\theta}}) - F^{-1}(\hat{\theta}_0). \tag{39}$$

Thus, we can transform uniquely between lengths $L_\theta$ and $L_{\hat{\theta}}$ so that the $\theta$ length constraint will be satisfied if and only if the $\hat{\theta}$ length constraint is satisfied.

In fact, because the $\hat{\theta}$ length gives a constraint on the integrated root-Fisher information, it can be transformed into stimulus space using $\sqrt{I_s}\mathrm{d}s = \sqrt{I_\theta}\mathrm{d}\theta$, to give:

$$L_{\hat{\theta}} = \int_{\theta_0}^{\theta_1} \sqrt{I_\theta(\theta)}\mathrm{d}\theta = \int_S \sqrt{I_s(s)}\mathrm{d}s. \tag{40}$$

This is the total root-Fisher constraint of earlier works [4, 5, 6] and, because it depends on the stimulus Fisher information, it is independent of how we choose to parameterize the problem.

### 3.4 Flat-Fisher distribution depends only on observable quantities

Here we derive the relationship between the flat-Fisher parameter distribution $p_{\hat{\theta}}(\hat{\theta})$ and two observable (or controllable) quantities: the stimulus distribution $p_s(s)$ and the Fisher information in the neural activity about the stimulus $I_s(s)$. The flat-Fisher parameter $\hat{\theta}$ is a monotonic increasing function of the parameter $\theta$, and thus a monotonic increasing function of the stimulus $s$. So,

$$p_{\hat{\theta}}\left(\hat{\theta}(s)\right) = p_s(s) \Big/ \frac{\mathrm{d}\hat{\theta}}{\mathrm{d}s}. \tag{41}$$

Because $\mathrm{d}\hat\theta = a\sqrt{I_\theta(\theta)}\mathrm{d}\theta$ and $I_s(s) = I_\theta(\theta(s))\frac{\mathrm{d}\theta}{\mathrm{d}s}^2$, we have

$$\frac{d\hat\theta}{ds} = \frac{d\hat\theta}{d\theta}\frac{d\theta}{ds} = a\sqrt{I_\theta(\theta(s))}\frac{\sqrt{I_s(s)}}{\sqrt{I_\theta(\theta(s))}}. \tag{42}$$

Hence,

$$\boxed{p_{\hat\theta}\left(\hat\theta(s)\right) = \frac{1}{a}\frac{p_s(s)}{\sqrt{I_s(s)}}.} \tag{43}$$

This can also be derived heuristically by taking the ratio of two changes of variable:

$$p_s(s)ds = p_\theta(\theta)d\theta = p_{\hat\theta}(\hat\theta)d\hat\theta, \tag{44}$$

$$\sqrt{I_s(s)}ds = \sqrt{I_\theta(\theta)}d\theta = \sqrt{I_{\hat\theta}(\hat\theta)}d\hat\theta = \frac{1}{a}d\hat\theta, \tag{45}$$

to get:

$$\frac{p_s(s)}{\sqrt{I_s(s)}} = \frac{p_\theta(\theta)}{\sqrt{I_\theta(\theta)}} = a p_{\hat\theta}(\hat\theta). \tag{46}$$

Thus,

$$p_{\hat\theta}(\hat\theta(s)) \propto \frac{p_\theta(\theta)}{\sqrt{I_\theta(\theta)}} = \frac{p_s(s)}{\sqrt{I_s(s)}}. \tag{47}$$

## 4 Adaptation of optimal codes to stimulus distribution

### 4.1 Stimulus distribution invariance implies log objective function

It is difficult to analyze the role of the stimulus distribution in the general efficient coding solutions that we have presented so far (Eqs. (18)-(20)) because the stimulus distribution comes into play in two places: in the argument to the objective function and in the $\gamma$ dynamics. Here we instead use a different approach to solving the optimization problem. Specifically, we reparameterize the parameter $\theta$ to the flat-Fisher space $\hat\theta$ *and* we reparameterize the stimulus to the space $\hat s = \mathrm{CDF}_s(s)$ in which the $\hat s$ probability is uniform. This means that stimulus density terms become $p_s(\mathrm{CDF}_s^{-1}(\hat s)) = \pi_s(\hat s)$ and that

$$\frac{\mathrm{d}\hat\theta}{\mathrm{d}\hat s} = \frac{1}{p_{\hat\theta}(\hat\theta(\hat s))}. \tag{48}$$

Now we solve the optimization by optimal control (see also Sec. (1.3)) keeping $\hat s$ as the independent variable. Denoting $u = \frac{\mathrm{d}\hat\theta}{\mathrm{d}\hat s}$ gives the Hamiltonian

$$H\left(\hat s, \hat\theta(\hat s), q(\hat s), u(\hat s)\right) = q(\hat s)u(\hat s) + f\left(u(\hat s)\pi_s(\hat s)\right) + \lambda C_{\hat\theta}\left(\hat\theta(\hat s)\right). \tag{49}$$

Applying the Pontryagin maximality principle (see Sec.(1.3) and [1, 3]) and rearranging the outputs gives

$$\boxed{\lambda C'_{\hat\theta}\left(\hat\theta(\hat s)\right) = \frac{\mathrm{d}}{\mathrm{d}\hat s}f'\left(u(\hat s)\pi_s(\hat s)\right)\pi_s(\hat s),} \tag{50}$$

a second order differential equation for $\hat\theta(\hat s)$.

Let us first examine the constraint free case. Setting $C'_{\hat\theta}(\hat\theta) = 0$, the differential Equation (50) can be integrated to give $f'(u\pi_s)\pi_s = \mathrm{const}$. Thus, the $\pi_s$ dependence will only cancel if $f'(x) \propto 1/x$, in which case a uniform solution (unconstrained, log objective) is achieved. Otherwise $u(\hat s)$ will be a non-trivial function of $\pi_s(\hat s)$ and stimulus distribution dependence will be assured. Thus, when constraints are absent, a log objective function is required for stimulus distribution independence.

In the presence of constraints, we can make a similar argument on the full differential Equation (50). Here the dependence on $\pi_s(\hat s)$ will again cancel if and only if $f$ is a logarithm so that $f' \propto 1/x$. Otherwise, the differential equation that determines $u(\hat s) = \mathrm{d}\hat\theta/\mathrm{d}\hat s = 1/p_{\hat\theta}(\hat\theta(\hat s))$ will depend on the probability density of the stimulus. This means that each stimulus distribution, $\pi_{si}$, will have

7

its own differential equation that determines the solution. Some of these differential equations will share solutions, for example stimulus distributions that are shifted from each other by a constant stimulus offset. We need to show that, unless the $\pi_s$ dependence explicitly cancels, not *all* of the different $\pi_{si}$ differential equations will share the *same* solution. This may be fairly obvious, but we give an argument here that also foreshadows the results of the next section.

Suppressing the $\hat{s}$ function arguments, assume that for a given stimulus distribution, $\pi_{s1}$, $u_1$ is the solution to the differential equation:

$$\lambda_1 C'_{\hat{\theta}}\left(\hat{\theta}_1\right) = \frac{\mathrm{d}}{\mathrm{d}\hat{s}} f'\left(u_1 \pi_{s1}\right) \pi_{s1}. \tag{51}$$

Because $\hat{\theta}(\hat{s})$ and its inverse map one bounded space to another, the function $1/u_1(\hat{s})$ can be normalized to a probability density and we can probe the optimal solution using this probability density $\pi_{s2} = A/u_1$ ($A$ constant). If the parameter distribution is adaptation invariant, the two stimulus distributions $\pi_{s1}$ and $\pi_{s2}$ will have the same solution: $u_1 = u_2$. This requires

$$\frac{\mathrm{d}}{\mathrm{d}\hat{s}} f'\left(u_1 \pi_{s1}\right) \pi_{s1} = \frac{\mathrm{d}}{\mathrm{d}\hat{s}} f'\left(u_2 \frac{A}{u_1}\right) \frac{A}{u_1} = f'(A) \frac{\mathrm{d}}{\mathrm{d}\hat{s}} \frac{A}{u_1}. \tag{52}$$

Thus, invariance requires

$$\frac{\mathrm{d}}{\mathrm{d}\hat{s}} f'\left(u_1 \pi_{s1}\right) \pi_{s1} \propto \frac{\mathrm{d}}{\mathrm{d}\hat{s}} \frac{1}{u_1}, \tag{53}$$

for all probability distributions $\pi_{s1}$. Integrating both sides, we see that this is only possible if (i) $f'(x) \propto 1/x$ and so $f$ is log, or (ii) $1/u_1 \propto \pi_{s1}$. In this second case, we have a 'fixed point' of the mapping because the distribution of parameters is proportional to the stimulus distribution, this is examined more below. For our current purpose, this proportionality must hold for all stimulus distributions $\pi_s$. However, this evidently requires that different stimulus distribution result in different functions $u$ and so cannot lead to stimulus distribution invariance. Thus, regardless of constraints, stimulus distribution invariance will only hold if the objective function is logarithmic.

### 4.2 Derivation of the adaptation fixed point

We have derived both the general solution to the efficient coding optimization and the transformation to the flat-Fisher space. We can now find the fixed point of neural adaptation in the flat-Fisher space. First, we rewrite the optimization solutions, Eqs. (18)-(20) in the flat-Fisher space, making use of Eq. (34),

$$\hat{f}\left(\frac{\pi_s\left(\hat{\alpha}\left(\hat{\theta}\right)\right)}{p_{\hat{\theta}}\left(\hat{\theta}\right)}\right) = \lambda C_{\hat{\theta}}\left(\hat{\theta}\right) + Z - \gamma\left(\hat{\theta}\right), \tag{54}$$

$$\frac{\mathrm{d}\gamma}{\mathrm{d}\hat{\theta}} = f'\left(\frac{\pi_s\left(\hat{\alpha}\left(\hat{\theta}\right)\right)}{p_{\hat{\theta}}\left(\hat{\theta}\right)}\right) \pi'_s\left(\hat{\alpha}\left(\hat{\theta}\right)\right), \tag{55}$$

$$\frac{\mathrm{d}\hat{\alpha}}{\mathrm{d}\hat{\theta}} = p_{\hat{\theta}}\left(\hat{\theta}\right). \tag{56}$$

Here $\hat{\alpha}$ is now the cumulative distribution function of $p_{\hat{\theta}}$, while the stimulus distribution dependent term $\pi_s(\cdot)$ remains unchanged.

We search for a fixed point in the sense of distributions: after adapting to a particular distribution of stimuli, $p_s$, the neural code will be characterized by a distribution of flat-space parameters, $p_{\hat{\theta}}$. When the 'input' stimulus distribution is the same as the 'output' parameter distribution, this can be said to be an adaptation fixed point. That is,

$$p_{\hat{\theta}}\left(\hat{\theta}\right) \propto p_s\left(s\left(\hat{\theta}\right)\right) = \pi_s\left(\alpha\left(\hat{\theta}\right)\right). \tag{57}$$

Proportionality allows the stimulus domain and parameter domain to have different lengths and, as we saw above, $p_{\hat{\theta}}(\hat{\theta})$ is only defined up to a proportionality constant due to rescaling of the $\hat{\theta}$ domain.

8

It may seem strange to compare distributions of stimuli to distributions of activity parameters, but it can be seen by integrating $p_s(s)ds = p_{\hat{\theta}}(\hat{\theta})d\hat{\theta}$ that, at such a fixed point,

$$\hat{\theta}(s) = as + b, \tag{58}$$

with $a$ and $b$ constants. Hence, $\hat{\theta}$ and $s$ are related through a linear scaling and shifting. Because $\hat{\theta}$ is only defined up to such a scaling and shifting, this is as close as we can come to identifying stimuli and (flat space) parameters. Intuitively, the distribution $p_{\hat{\theta}}$ is shaped by two factors: (i) the likelihood of the stimulus that each value of $\hat{\theta}$ encodes, and (ii) the degree to which the codes for nearby stimuli are stretched apart or compressed together in parameter space. At the fixed point, there will be no stretching or compression, so that the parameter probability density is determined exclusively by the probability density of the stimulus encoded.

At such a fixed point, Equation (57) shows that the argument to $\hat{f}$ in Eq. (54), and the argument to $f'$ in Eq. (55) become constant,

$$\frac{\pi_s\left(\alpha\left(\hat{\theta}\right)\right)}{p_{\hat{\theta}}\left(\hat{\theta}\right)} = c_0. \tag{59}$$

Taking Eq. (55) and multiplying the right-hand side of by $c_0$ and also dividing it by the equivalent ratio of probability densities from Eq. (59) gives:

$$\frac{d\gamma}{d\theta} = f'(c_0)\pi'_s\left(\alpha\left(\hat{\theta}\right)\right)c_0\frac{p_{\hat{\theta}}\left(\hat{\theta}\right)}{\pi_s\left(\alpha\left(\hat{\theta}\right)\right)} = c_1\frac{d}{d\theta}\log\pi_s\left(\alpha\left(\hat{\theta}\right)\right) \tag{60}$$

This is a constant multiple of the log-objective function solution in Eq. (21), and can be substituted into Eq. (54) to give (combining additive constants into $Z$ and multiplicative constants in $\lambda$ and $Z$)

$$\log\pi_s\left(\alpha\left(\hat{\theta}\right)\right) = \lambda C_{\hat{\theta}}\left(\hat{\theta}\right) + Z. \tag{61}$$

Hence,

$$p_{\hat{\theta}}\left(\hat{\theta}\right) \propto \pi_s\left(\alpha\left(\hat{\theta}\right)\right) \propto \exp\left(\lambda C_{\hat{\theta}}\left(\hat{\theta}\right)\right). \tag{62}$$

Because it is a function of the parameter directly, the constraint function, $C_{\hat{\theta}}(\hat{\theta})$ will be dependent on the particular parameterization that is used, here the flat-Fisher space parameterization. If required, it can be converted back to $\theta$-space given knowledge of $I_\theta(\theta)$ via the integrated transformation from above, $C(\theta) = C_{\hat{\theta}}(F(\theta))$.

Once we know $C_{\hat{\theta}}(\hat{\theta})$, we can also recover the objective function. Probing with a uniform stimulus distribution, we know that the resulting flat-space distribution follows

$$p_{\hat{\theta}}(\hat{\theta}) = \frac{p_s}{\hat{f}^{-1}(\lambda C_{\hat{\theta}}(\hat{\theta}) + Z)}. \tag{63}$$

This allow us to fit $\hat{f}^{-1}$ by regressing $C_{\hat{\theta}}(\hat{\theta}(s))$ against $1/p_{\hat{\theta}}(\hat{\theta}(s))$. This does require determining the function $\hat{\theta}(s)$ for the code adapted to the uniform stimulus distribution, which can be accomplished by integrating (numerically) $1/p_{\hat{\theta}}(\hat{\theta}(s))ds$.

## 5 Estimating the distribution of neural activity parameters from data

Because the distribution $p_{\hat{\theta}}(\hat{\theta}(s))$ is crucial for characterization of the objective and constraint functions that shape a neural code, we showed in a main text a proof of principle for recovery of $p_{\hat{\theta}}(\hat{\theta}(s))$ from neural activity data. Here we present the details of the model architecture and training used.

### 5.1 Parameter regression

The first step in recovery of the flat space parameter distribution from activity data, $\{s_i, \boldsymbol{r}_i\}_i$, is regression from stimuli to the parameters underlying the neural activity. Given a known or identified

9

neural activity model, $p(\boldsymbol{r}|\boldsymbol{\eta}(s)) = \exp(\boldsymbol{\eta}(s)^T \boldsymbol{T}(\boldsymbol{r}) - A(\boldsymbol{\eta}(s)))$, we aim to find the function $\boldsymbol{\eta}(s)$ by regression. Here, we use a neural network to perform maximum likelihood regression. Parameterizing $\boldsymbol{\eta}(s)$ by weights, $\boldsymbol{w}$, we can compute gradients of the log likelihood of the data under the model according to:

$$\frac{\partial L}{\partial \boldsymbol{w}} = \left( \sum_i \boldsymbol{T}(\boldsymbol{r}_i) - \frac{\partial A(\eta(s_i))}{\partial \boldsymbol{\eta}} \right) \frac{\partial \boldsymbol{\eta}}{\partial \boldsymbol{w}}, \tag{64}$$

where $L$ is the log-likelihood of the activity data, $L = \sum_i \log p\left(\boldsymbol{r}_i | \boldsymbol{\eta}(s_i)\right)$. Here $\boldsymbol{T}(\boldsymbol{r})$ are the sufficient statistics and $\partial A / \partial \boldsymbol{\eta}$ is the parameter to moment mapping of our activity model. Concretely, the term in parentheses is calculated from the data and model output, and used as the initial gradient for backpropagation to produce the weight gradients.

For the example in the main text, we used a neural network with 4 layers of width 3 and tanh nonlinearity implemented in PyTorch. We trained on 2k data points with 20k repeats of the training set and mini-batches of size 100 using the Adam optimizer with learning rate, $10^{-3}$ and regularization by weight decay with strength $10^{-2}$. Training hyperparameters and network architecture were set manually based on recovery of a validation $\boldsymbol{\eta}(s)$ curve (different from the test curve) under the same (Poisson noise) loss function. Hyperparameters were probed within a 10-fold range for training hyperparameters and a 2-fold range for architecture parameters.

### 5.2 Extracting the flat-Fisher distribution

Having performed the parameter regression, we now want to estimate the parameter distribution in the flat-Fisher space. This can be accomplished using Eq. (43),

$$p_{\hat{\theta}}\left(\hat{\theta}(s)\right) \propto \frac{p_s(s)}{\sqrt{I_s(s)}}. \tag{65}$$

We know $p_s(s)$, so all that is left is to find the stimulus Fisher information, $I_s(s)$. Here, we use the Fisher information in our trained model of neural activity as a proxy for the stimulus Fisher information in the neural activity. With the parameter mapping $\boldsymbol{\eta}(s)$ from our trained model, this is given by:

$$I_s(s) = \frac{d\boldsymbol{\eta}}{ds}^T I_{\boldsymbol{\eta}}(\boldsymbol{\eta}(s)) \frac{d\boldsymbol{\eta}}{ds}. \tag{66}$$

The $\boldsymbol{\eta}$ derivatives can be computed by forward differentiation of the fit $\boldsymbol{\eta}(s)$ function. The model Fisher information function, $I_{\boldsymbol{\eta}}(\cdot)$, is specified by our exponential family model of the noise in neural activity, we take the function to be known and evauluated on fit outputs $\boldsymbol{\eta}(s)$. Alternatively, it could be sampled, using the identity $I_{\boldsymbol{\eta}}(\boldsymbol{\eta}) = \mathrm{cov}_{p(\boldsymbol{r}|\boldsymbol{\eta})}[\boldsymbol{T}(\boldsymbol{r})]$.

Instead of computing $I_s$ and dividing the result out from the (known) stimulus distribution, we can also CDF transform the incoming data prior to training, $\tilde{s} = \mathrm{CDF}_s(s)$ so that the output parameters are $\boldsymbol{\eta}(\tilde{s})$. In this case, the parameter derivatives are:

$$\frac{d\boldsymbol{\eta}}{d\tilde{s}} = \frac{d\boldsymbol{\eta}}{ds}\frac{ds}{d\tilde{s}} = \frac{d\boldsymbol{\eta}}{ds}\frac{1}{p_s(s(\tilde{s}))}. \tag{67}$$

Then,

$$I_{\tilde{s}} = \frac{d\boldsymbol{\eta}}{d\tilde{s}}^T I_{\boldsymbol{\eta}}(\boldsymbol{\eta}(\tilde{s}))\frac{d\boldsymbol{\eta}}{d\tilde{s}} = \frac{d\boldsymbol{\eta}}{ds}^T I_{\boldsymbol{\eta}}(\boldsymbol{\eta}(\tilde{s}))\frac{d\boldsymbol{\eta}}{ds}\frac{1}{p_s(s(\tilde{s}))^2} = \frac{I_s(s(\tilde{s}))}{p_s(s(\tilde{s}))^2}. \tag{68}$$

Hence,

$$p_{\hat{\theta}}(\hat{\theta}(s)) \propto \frac{1}{\sqrt{I_{\tilde{s}}(\mathrm{CDF}_s(s))}}. \tag{69}$$

The constant of proportionality can be found by normalization of this function.

### References

[1] Daniel Liberzon. *Calculus of variations and optimal control theory: a concise introduction.* Princeton University Press, 2011.



\end{document}